\documentclass[preprint,12pt]{elsarticle}
\makeatletter
\def\ps@pprintTitle{%
 \let\@oddhead\@empty
 \let\@evenhead\@empty
 \def\@oddfoot{\centerline{\thepage}}%
 \let\@evenfoot\@oddfoot}
\makeatother

\usepackage{amssymb}
\usepackage{graphicx}
\usepackage{amsmath}
\usepackage[labelformat=simple]{subfig} 
\usepackage{xcolor}

\begin{document}

\begin{frontmatter}



\title{Fighting together against the pandemic: learning multiple models on tomography images for COVID-19 diagnosis}



\author{Mario Manzo}
\address{IT Services, University of Naples “L’Orientale”\\
              80121, Naples, Italy, mmanzo@unior.it}
\author{Simone Pellino}
\address{Department of Applied Science,\\ I.S. Mattei Aversa 81031,  M.I.U.R. Rome, Italy; simonepellino@gmail.com}



\begin{abstract}
The great challenge for the humanity of the year 2020 is the fight against COVID-19. The whole world is making a huge effort to find an effective vaccine with purpose to protect people not yet infected. The alternative solution remains early diagnosis, carried out through real-time polymerase chain reaction (RT-PCR) test or thorax computer tomography (CT) scan images. Deep learning algorithms, specifically convolutional neural networks, represent a methodology for the image analysis. They optimize the classification design task, essential for an automatic approach on different types of images, including medical. In this paper, we adopt pretrained deep convolutional neural network architectures in order to diagnose COVID-19 disease on CT images. Our idea is inspired by what the whole of humanity is achieving, substantially the set of multiple contributions is better than the single one for the fight against the pandemic. Firstly, we adapt, and subsequently retrain, for our assumption some neural architectures adopted in other application domains. Secondly, we combine the knowledge extracted from images by neural architectures in an ensemble classification context. Experimental phase is performed on CT images dataset and results obtained show the effectiveness of the proposed approach with respect to state-of-the-art competitors.\end{abstract}




\end{frontmatter}


\section{Introduction}
\label{intro}

The proliferation of the new coronavirus, from now COVID-19, is the current threat to humanity and has spread rapidly around the world starting from January 2020. The 30th of January 2020 is a reference for history because it has been declared by the World Health Organization (WHO) as the official start of international public health emergency, better known as a pandemic. Currently, no countries is immune to the virus and, clearly, the situation appears to be critical. The virus manifests itself after 5-6 days of the onset of the disease with specific and non-specific symptoms. The first are fever, dry cough, sore throat, loss of taste or smell, while the second are fatigue, headache and breathlessness. Unfortunately, COVID-19 has also occurred in animals through transmission between them. Looking back, viruses with similar behavior are Severe Acute Respiratory Syndrome Coronavirus (SARS-CoV) and the Middle East Respiratory Syndrome Coronavirus (MERS Coronavirus) with related major respiratory problems. Currently, the medical protocol takes more than 24 hours to detect the virus in the human body. It is important to detect the disease during the starting phase in order to isolate the infected person because there is no effective cure. Diagnosis can be made through real-time polymerase chain reaction (RT-PCR). RT-PCR is not very reliable due to the high false negative rates and finalization time. Otherwise, COVID-19 can be detected in healthy people due to a false positive. It is clear that the low sensitivity of RT-PCR test is not satisfactory in the current pandemic situation. In some cases, the infected are not recognized in time and do not receive adequate care. As an alternative to RT-PCR, the Thorax Computer Tomography (CT) is probably a more reliable, effective, and faster approach for virus detection and treatment. In almost all hospitals CT image screening is available and can be adopted for a first analysis of the virus. Unfortunately, the Thorax CT images requires a radiologist and a lot precious time is lost. Therefore, the automated analysis of Thorax CT images can speed up the diagnosis in order to help specialist medical staff, above all, and not only, to avoid delays in the start of treatment. In the last few years, deep learning has proven effective for the management, analysis, representation and classification of medical images. In particular, the success of deep neural networks, applied to the image classification task, is connected to different interesting aspects such as the spread of software in terms of open source license, the constant growth of hardware power and the availability of large datasets. Specifically, for the treatment of COVID-19, deep neural networks are adopted both in segmentation and detection phases. However, uncertainty in COVID-19 diagnosis and data imbalance have a decisive impact on performance, hampering model generalization. In order to provide a solution to the above issues, we introduce a framework based on transfer deep learning and ensemble classification for COVID-19 diagnosis. It works based on three integrated stages. A first, which performs image preprocessing operations such as image resize and augmentation. A second, which redesigns and retrains multiple deep neural networks. A third, which combines different predictions provided by deep neural networks with the aim of making the best decision (COVID-19/not-COVID-19). The framework provides the following main contributions:
\begin{itemize}
\item A deep and ensemble learning based framework, to simultaneously address variation between classes and class imbalance for COVID-19 diagnosis task.
\item A framework that provides multiple classification models, based on deep transfer learning.
\item The demonstration that choosing multiple models, suitably combined, is better than a single model and can strengthen the decision during the diagnosis by a specialist doctor.
\item Some experimental greater improvements over existing methods on recent state of art dataset about COVID-19 detection task.
\end{itemize}

The paper is structured as follows. Section \ref{relatedwork} provides an overview of state of art about COVID-19 classification approaches. Section \ref{MM} describes in detail proposed framework. Section \ref{res} provides a wide experimental phase, while section \ref{conc} concludes the paper.

\section{Related work}
\label{relatedwork}
In this section, we briefly analyze the most important approaches working on COVID-19 diagnosis now existing in literature. In this field are included numerous works that address the task according to different aspects. Some offer important contributions about image representation, by implementing segmentation algorithms or new descriptors. Instead, others implement complex mechanisms of learning and classification. 

In \cite{loey2020deep} authors propose an architecture in order to improve performance in recognizing COVID-19 from chest radiograph images. It consists of two main components: image augmentation and transfer learning. This combination improves performance measurements such as accuracy, sensitivity, specificity, precision, accuracy, and F1 score. 

Authors in \cite{amyar2020multi} present a multitask deep learning model to jointly identify COVID-19 patient and segment COVID-19 lesion from chest computed tomography images. The proposed architecture includes three phases: COVID-19 vs normal vs other infections classification, COVID-19 lesion segmentation, image reconstruction. Furthermore, algorithm flow is based on a common created encoder for the three tasks. It takes a CT scan as input, and its output is then adopted for image reconstruction via a first decoder, to the segmentation via a second decoder, and to the classification of COVID-19 vs normal vs other infections via multilayer perceptron.

In \cite{he2020sample} authors build a public available dataset containing hundreds of CT scans COVID-19 positive and implement a sample-efficient deep learning approach that can obtain high diagnosis accuracy on a limited training set CT images. The approach integrates contrastive self-supervised learning with transfer learning layer, to learn powerful and unbiased features representation. Aim to reducing the risk of overfitting, a large and consistent dictionary on-the-fly based on the contrastive loss to fulfill this auxiliary task is built.

In \cite{el2020novel} authors propose a features selection and voting classifier framework for COVID-19 CT image classification. Firstly, the features are extracted using convolutional neural network (AlexNet). Secondly, a proposed features selection algorithm, Guided Whale Optimization Algorithm (SFS-Guided WOA) based on Stochastic Fractal Search (SFS), is then applied followed by balancing algorithm. Finally, a voting approach, Guided WOA based on Particle Swarm Optimization (PSO), that aggregates different classifier such as Support Vector Machine (SVM), neural networks, K-Nearest Neighbor (KNN) and decision trees predictions, to choose the most voted class in an ensemble learning way, is adopted.

In \cite{shah2020diagnosis} authors design a neural architecture, called CTnet-10, for the COVID-19 diagnosis from CT images. It is formed by a max-pooling layer of dimension $62\times62\times32$ followed by 2 convolutional layers of dimensions $60\times60\times32$, $58\times58\times32$ respectively and a pooling layer of dimension $29\times29\times32$. The last levels are: a flattened layer which is connected out to an fully connected layer of 4096 neurons, in which the dropout layer was used in each of these. Last layer, a single neuron sigmoid and linear, classifies CT scan images as COVID-19 positive or negative. Tests results are compared with known neural networks (DenseNet-169, VGG-16, ResNet-50, InceptionV3, and VGG-19).

In \cite{zhao2020relation} authors build an open-source COVID-19 CT image dataset and a diagnosis method based on multi-task learning and self-supervised learning. To address the overfitting issue, they study two strategies: one is to add additional information including segmentation masks of lung regions and fed them into the feature extraction network.

In \cite{bernheim2020chest} authors does a retrospective study on chest CT scans images with purpose to find the relationship to the time between symptom onset and the initial CT scan. The hallmarks of COVID-19 infection on images are bilateral and peripheral ground-glass and consolidative pulmonary opacities. With a longer time after the onset of symptoms, CT findings are more frequent, including consolidation, bilateral and peripheral disease, greater total lung involvement, linear opacities, crazy-paving pattern, and the reverse halo sign.

In \cite{gozes2020rapid} authors develop an AI-based automated CT image analysis tools for detection, quantification, and tracking of COVID-19. The system utilizes robust 2D and 3D deep learning models, modifying and adapting existing AI models and combining them with clinical understanding. The first step is the lung crop stage, in which the lung region of interest is extracted using a lung segmentation module. The following step detects COVID-19 related abnormalities using deep convolutional neural network architecture. To overcome the limited amount of images dataset, data augmentation techniques (image rotations, horizontal flips and cropping) are applied.

In \cite{zheng2020deep} author propose a 3D deep convolutional neural network, named DeCoVNet, to detect COVID-19 from CT volumes. DeCoVNet is composed of three blocks. First, called network stem, which consists in a vanilla 3D convolution, a batchnorm layer and a pooling layer. Second, composed of two 3D residual blocks (ResBlocks). In each ResBlock, a 3D feature map is passed into both a 3D convolution with a batchnorm layer and a shortcut connection containing a 3D
convolution. The output feature maps are added in an element-wise manner. Third, a progressive classifier (ProClf), which is composed of three 3D convolution layers and a fully-connected (FC) layer. A softmax activation function progressively abstracts the information in the CT volumes by 3D max-pooling and finally directly outputs the probabilities of being COVID-19 or not. 

In \cite{ai2020correlation} authors investigate the diagnostic value and consistency of chest CT as comparison with RT-PCR test. For patients with multiple RT-PCR, the dynamic conversion of RT-PCR results (negative to positive, positive to negative) is analyzed as compared with serial chest CT scans for those with a time interval between RT-PCR tests of 4 days or more. Chest CT has a high sensitivity for diagnosis of COVID-19 and may be considered as a primary tool for the current detection in epidemic areas.

In \cite{fang2020sensitivity} authors examine the sensitivity, specificity, and feasibility of chest CT in detecting COVID-19 compared with RT-PCR test. Sensitivity and specificity of chest CT in their various steps are compared using RT-PCR as a gold standard. A reverse calculation approach is applied to chest CT as a hypothetical gold standard and compared to RT-PCR to it point out the flaw of the standard approach. The study want to prove that the sensitivity and specificity of the chest CT in COVID-19 diagnosis and the radiation exposure have to be taken into account together.

\section{Materials and Methods}
\label{MM}
In this section we introduce the proposed framework which is composed of two well known methodologies: deep neural networks and ensemble learning. The main idea is to combine several deep neural networks with purpose to classify images. The result is a set of competitive models providing a range of confidential decisions useful for making choices during classification. The framework is structured into three level. A first, which performs preprocessing in terms of image resize and augmentation. A second, which learns different deep neural networks, previously redesigned for the specific task. A third, in which different models provided by deep neural networks are combined, through ensemble rules, for classification purpose. Finally, the framework iterates through a predetermined number of times in a supervised learning context.

\subsection{Image augmentation}
\label{imgaug}

Many approaches have been developed to address the complications associated with the limited amount of data in machine learning. Image augmentation \cite{shorten2019survey} is a functional technique for increasing and/or changing the size of the training set without acquiring additional images. The concept is basic and consists of duplicating and/or modifying the images with some kind of variation so that more samples can help to train the model. As general idea, the image is augmented in a way that preserves key features for making predictions, but reworked so that the pixels present in some form of noise. The augmentation will be harmful if produces images that are very dissimilar to those used to test the model, so it is clear that this process must be organized in detail. In the proposed framework, we have adopted random reflection, translation, and scaling in order to enhance and augment the image content. As described in the experimental section, this step turns out to be fundamental in improving the performance of the proposed approach.

\subsection{Image resize}
\label{resize}

One of the defects of neural networks concerns the fixed size of the images to be processed. To this end, a resize step is performed based on the input layer dimension claimed by the deep neural networks (details can be found in table \ref{nets} column 5). Most of the networks need this trick but it does not alter the image information content in any way. The size normalization is essential because images of different or large dimensions cannot be processed for the network training and classification stages.


\subsection{Network design and transfer learning}
\label{netmod}

The transfer learning approach has been selected for classification purpose. The basic idea is to transfer the knowledge extracted from a source domain to a destination one, in our case the COVID-19 diagnosis. Generally, a pretrained network is chosen as starting point in order to learn a new task. It is the easiest and fastest solution to adopt the representational power of pretrained deep networks. Clearly, it turns out to be much faster and easier to tune a network with transfer learning than training a new network from scratch with randomly initialized weights. For COVID-19 diagnosis, deep learning architectures are selected based on their structure and performance skills. The goal is to train networks on images by redesign their structures in the final layer according to the needs of the addressed task (two outgoing classes: COVID-19 and not-COVID-19). Table \ref{nets} supports the provided below description of adopted networks.

Resnet18 \cite{Resnet} is inspired by pyramidal cells contained in the cerebral cortex. It uses particular skip connections or shortcuts to jump over some layers. It is composed of 18 layers deep, which with the help of a technique known as skip connection has paved the way for residual networks.

Densenet201 \cite{huang2017densely} is a convolutional neural network with 201 layers deep. Unlike standard convolutional networks composed of $L$ layers with $L$ one-to-one connections between the current layers and the nexts, it contains $\frac{L(L+1)}{2}$ direct connections. Specifically, each layer adopts the feature-maps of all preceding layers and its own feature-maps into all subsequent layers as inputs.

Mobilenetv2 \cite{sandler2018mobilenetv2} is a convolutional neural network with 53 layers deep. It was built based on inverted residual structure with shortcut connections between the thin bottleneck layers. The intermediate expansion layer uses lightweight depthwise convolutions to filter features as a source of non-linearity. Furthermore, non-linearities in the narrow layers are removed with purpose to maintain representational power.

Shufflenet \cite{zhang2018shufflenet} is a convolutional neural network with 173 layers deep designed for mobile devices with very limited computing power. The peculiarity of this architecture concerns the introduction of pointwise group convolution and channel shuffle operations in order reduce computation cost and maintain accuracy.

\begin{table}[!ht]
\centering \caption{Description of adopted pretrained network.}
\footnotesize
\begin{tabular}{l l l l l}
\hline
  Network & Depth & Size (MB) & Parameters (Millions) & Input Size\\ \hline
  Resnet18 & 18 & 44 & 11.7 & 224 $\times$ 224\\ \hline
  Densenet201 & 201 & 77 & 20.0 & 224 $\times$ 224\\ \hline
  Mobilenetv2 & 53 & 13 & 3.5 & 224 $\times$ 224\\ \hline
  Shufflenet & 50 & 6.3 & 1.4 & 224 $\times$ 224\\ \hline
\end{tabular}
\label{nets}
\end{table}

Deep neural networks have been adapted to COVID-19 classification problem. Originally, the main training phase is performed on the Imagenet dataset \cite{deng2009imagenet}, which include a million images divided into 1000 classes. The result consists in a rich features representation for a wide range of images. The network processes an image and provides a prediction about a classes to which it could belong with an attached probability. Commonly, the first layer of the network is the image input layer. The input requires images with 3 color channels. Immediately after is followed by convolutional layers, which work with purpose to extract image features. Particularly, last learnable layer and the final classification layer are adopted to classify the input image. In order to make the pretrained network compliant to classify new images, the two last layers with new layers are replaced. Frequently, the last layer, with related learnable weights, is fully connected. This layer is removed and replaced by a new which is fully connected with outputs related to number of classes of new data (COVID-19 and not-COVID-19). In addition, learning phase of the new layer, compared to the transferred layers, can be speeded up by increasing the rate factors. Optionally, the weights of previous levels can be left unchanged by setting their learning rate to zero. This variation prevents the updating of the weights, during training, and a consequent lowering of the execution time as the gradients of the relative layers do not have to be calculated. This aspect has a strong impact in the case of small datasets in order to avoid overfitting.





\subsection{Ensemble Learning}
\label{ense}

The contribution of different deep neural networks can be mixed in an ensemble context. Considering the set, with cardinality $k$, of images belonging to $x$ classes, to be classified

\begin{equation}
Imgs=\{i_{1},i_{2},\ldots,i_{k}\}    
\end{equation}

each element of the set will be treated with the procedure below. Let's consider the set $C$ composed of $n$ deep neural networks

\begin{equation}
C=\{\beta_{1},\beta_{2},\ldots,\beta_{n}\}    
\end{equation}




which are combined to classify the images through the set $CN$

\begin{equation}
CN = \begin{bmatrix} 
    \beta_{1}i_{1} & \dots & \beta_{1}i_{k} \\
    \vdots & \ddots & \\
    \beta_{n}i_{1} &        & \beta_{n}i_{k} 
    \end{bmatrix}
\end{equation}
each $\beta_{n}$ provides a decision $d \in I \{-1,1\}$, where $1$ stands for not-COVID and $-1$ for COVID, with reference to $i_k \in Imgs$. The set of decisions $D$ can be defined as follows


\begin{equation}
D = \begin{bmatrix} 
    d_{\beta_{1}i_{1}} & \dots & d_{\beta_{1}i_{k}} \\
    \vdots & \ddots & \\
    d_{\beta_{n}i_{1}} &        & d_{\beta_{n}i_{k}} 
    \end{bmatrix}
\end{equation}

it should be noted that each element of the matrix $D$ corresponds to the result of the deep neural network and image combination of the matrix $CN$ in terms of position, such as $\beta_{n}i_{k}\to d_{\beta_{n}i_{k}}$. Furthermore, a score value $s$, $s \in S \{0,\dots,1\}$, is associated with each decision $d$ and represents the posterior probability $P(i|x)$ that an image $i$ could belong to class $x$.  In addition, the set of scores $S$ can be defined as follows

\begin{equation}
S = \begin{bmatrix} 
    P(i_{1}|x)_{d_{\beta_{1}i_{1}}} & \dots & P(i_{k}|x)_{d_{\beta_{1}i_{k}}} \\
    \vdots & \ddots & \\
    P(i_{1}|x)_{d_{\beta_{n}i_{1}}} &        & P(i_{k}|x)_{d_{\beta_{n}i_{k}}} 
    \end{bmatrix}
\end{equation}

also in this case each element of the matrix $S$ corresponds to the result of the deep neural network and image combination of the matrix $CN$ with related posterior probability in terms of position, such as $\beta_{n}i_{k}\to d_{\beta_{n}i_{k}} \to P(i_{k}|x)_{d_{\beta_{n}i_{k}}}$. At this point, let's introduce the concept of mode, defined as the value which is repeatedly occurred in a given set

\begin{equation}
mode=l+\left(\frac{f_1-f_0}{2f_1-f_0-f_2}\right)
\times h
\end{equation}

where $l$ is the lower limit of the modal class, $h$ is the size of the class interval, $f_1$ is the frequency of the modal class, $f_0$ is the frequency of the class which precedes the modal class and $f_2$ is the frequency of the class which successes the modal class. The columns of matrix $D$ are analyzed by applying the mode, in order to obtain the values of the most frequent decisions. This step is performed in order to verify the best responses of the different deep neural networks, contained in the set $C$. Moreover, the meaning of mode is twofold. First, the most frequent value. Second, its occurrences in terms of indices. For each most frequent occurrence, modal value, the corresponding score from the matrix $S$ is extracted. In this regard, a new vector is generated

\begin{equation}
DS=\{ds_{P(i_{1}|x)_{d_{\beta_{1,\dots,n}i_{1}}}}, \ldots, ds_{P(i_{k}|x)_{d_{\beta_{1,\dots,n}i_{k}}}} \},\end{equation}

where each element $ds$ contains the average of the decision scores with higher frequency, extracted through the mode, in the related column of the matrix $D$. Also, the modal value of each column of the matrix $D$ is stored in the vector $DM$

\begin{equation}
DM=\{dm_{d_{\beta_{1,\dots,n}i_{1}}},\dots, dm_{d_{\beta_{1,\dots,n}i_{k}}}\},\end{equation}

each value $dm$ contains the modal value of the class to which image $i$ could belong with the average probability score $ds$. In essence, this is the class to which an image could belong based on the votes given by different deep neural networks.

\section{Experimental results}
\label{res}

This section describes the experiments performed on public dataset. In order to produce compliant performance, the settings reported in recent COVID-19 classification methods are adopted. The experimental phase is structured in two parts with purpose to address the COVID-19 detection task. The first concerns a comparison with deep neural networks, in order to prove how a multiple model can provide better guidance than a single. While, the second deals with recent methods, which adopt a different logic than the proposed approach.


\subsection{Dataset}
\label{dataset}

The COVID-19 CT adopted dataset is publicly available \footnote{https://github.com/UCSD-AI4H/COVID-CT} and the details are described in \cite{zhao2020covid}. It is composed of 746 Thorax Computer Tomography (CT) images, where 349 contain clinical findings of COVID-19 from 216 patients and 397 are obtained from non-COVID-19 patients. The CT images are a collection selected from COVID-19 related papers published in medRxiv, bioRxiv, NEJM, JAMA, Lancet, and others. The reliability of this dataset has been validated by a senior radiologist, of Tongji Hospital, Wuhan, China, that has worked on diagnosis and treatment of a large number of COVID-19 patients during the period of maximum emergency between January and April 2020.



\subsection{Settings}
\label{settings}

The framework consists in different modules written in Matlab language and extends the code available in \cite{pham2020comprehensive}. The pretrained networks were taken from ImageNet Large-Scale Visual Recognition Challenge (ILSVRC) \cite{russakovsky2015imagenet}. Various combinations of networks were selected, because the test phase was split by applying and not image augmentation. The choice is not random but was made based on single network performance and an in-depth analysis of their architecture (number of layers, applications in literature, etc). A different combination did not provide expected feedback. Densenet201, Mobilenetv2, Resnet18 were adopted with image augmentation, while Shufflenet, Resnet18 and Mobilenetv2 were adopted without image augmentation. Table \ref{nets} shows some important details related to the adopted networks. Among all the computational stages, the training process was certainly the most expensive. As is certainly known, the networks are composed of fully connected layers that make the structure extremely dense and complex. This aspect certainly increases the computational load. In order to compare the results with those obtained in \cite{pham2020comprehensive}, networks were trained by setting the mini batch size to $5$, the maximum epochs to $6$, the initial learning rate to $3 \cdot 10^{-4}$ (constant for the training stage), momentum value to $0.9$, gradient threshold method to L2 norm, factor for L2 regularization (weight decay) to $1 \cdot 10^{-4}$,  minimum batch size to $10$ and the optimizer is stochastic gradient descent with momentum (SGDM) algorithm. $80\%$ and $20\%$ of images are randomly included in train and test set respectively, for a number of iteration equal to $5$ with the aim of calculate the relevance feedback measures reported in table \ref{Metric}. For each training and before each network validation epoch the data were shuffled. The images were converted into RGB space and resized to align with the input format of each pretrained network.
The training was performed with and without image augmentation. Random reflection, translation and scaling were performed for the option of image augmentation. About random reflection, each image was reflected vertically with probability equal to $0.5$ in the top-bottom direction. Again, a horizontal and vertical translation to each image was applied. The translation distance was selected randomly from a continuous uniform distribution within the specified range $[-30,30]$. Similarly, images were scaled vertically and horizontally by selecting in random way the scale factor from a continuous uniform distribution within the specified range $[0.9, 1.1]$.




\subsection{Relevance feedback}
\label{RF}

Table \ref{Metric} summarizes the metrics adopted for the performance evaluation. The goal is to provide an uniform comparison with approaches working on the same task and to understand, from experimental phase, what information can be useful for COVID-19 diagnosis. 

\begin{table}[!ht]
\centering
\caption{Evalutation metrics adopted during relevance feedback stage.}
\begin{tabular}{c c}
 &      \\

\hline \textbf{Metric}   & \textbf{Equation}  \\

\hline Sensitivity         & $\frac{TP}{TP + FN}$                     \\

\hline  Specificity      & $\frac{TN}{TN + FP}$                    \\

\hline Accuracy         & $\frac{TP+FN}{TP + FP + TN + FN}$                     \\

\hline $F_{1}$         &$\frac{2 \cdot TP}{2 \cdot TP+FP+FN}$\\





\hline
 &                                                                      
\end{tabular}
\label{Metric}
\end{table}

The Sensitivity, also known as True Positive rate, concerns the portion of images containing COVID-19 disease elements that are correctly identified. The measure provides important information because highlights the skill to detect images containing the disease and contributes to increase the degree of robustness of result. At the same, it is possible to state of the Specificity, also known as True Negative rate, which instead measures the portion of negatives, images not containing COVID-19 disease elements, that have been correctly identified. Differently, accuracy, a well-known performance measure, is the proportion of true results among the total number of cases examined. In our case provides an overall analysis, certainly a rough measurement compared to the previous ones, about the skill of a classifier to distinguish an image of patient affected to COVID-19 from an image of patient not affected to COVID-19. Furthermore, $F_{1}$ is defined as the combination of precision and recall of the model in term of harmonic mean. In addition, the AUC was calculated using the trapezoidal integration to estimate the area under the ROC curve and represents the measure of performance of a classifier. ROC is a probability curve built by showing the True Positive rate against the False Positive rate with different threshold values. The AUC value is contained in the range between 0.5 and 1, where the value 0.5 represents the performance of a random classifier and the value 1 indicates a perfect one. A high AUC value provides positive classification indications. 

\subsection{Discussion}
\label{discussion}

Tables \ref{reswithau} and \ref{reswithoutau} describe the comparison with different deep neural networks respectively applying and not image augmentation. The provided performance can be considered satisfactory compared to different neural architectures. In terms of accuracy, although it provides a rough measurement, we have provided the best result with and without image augmentation. Sensitivity, a measure that provides greater confidence about addressed problem, is very high for both cases. Otherwise, Specificity, which also provides a high degree of information related to the absence of COVID-19 within the image, is the best value for both cases. Regarding the remaining measures, F$_{1}$ score and AUC, considerable values were obtained. Table \ref{comparison} provides comparison results with existing COVID-19 classification methods in term of accuracy. As shown, the proposed approach is only surpassed by Ai et al. \cite{ai2020correlation} and Fang et al. \cite{fang2020sensitivity}. For the remaining methods the performance provided is better. The effectiveness of the results can be attributed to two main aspects: deep neural networks and competitive model for classification. First, the deep neural networks chosen for images learning and classification are the main strong point. Furthermore, the framework provides multiple learning models that certainly constitute a different starting point than a standard approach, in which a single model is provided. This aspect is relevant for improving performance. Second, the classification stage which provides multiple choices in decision making. In fact, at each iteration, the framework selects which networks are suitable for recognizing COVID-19 in the images on test set. Certainly, the computational load is greater but produces better results than a single classification approach. Not negligible issue concerns the image size normalization, with respect to the request of the first layer of the neural networks, before the leaning phase, which not produce a performance degradation. In other cases, degradation of image details, quality and content is due to normalization. Otherwise, the weak point is the computational load even if pretrained networks include layers with already tuned weights. Surely, the time required for training stage is long and computational resources are high but less than a network created from scratch. Moreover, the addressed binary classification has not been greatly disadvantaged by the problem of class imbalance because the relationship between the number of images per class is not very unbalanced. In many cases, data imbalance impacts a low prediction of accuracy for the minority class. It is open problem but the solutions are many such as undersampling of majority class or oversampling of minority classes using image augmentation, weighted loss method by updating the loss function to result in the same way for all classes. This behavior is often seen in medical data due to the limitations of patient samples and cost of acquiring annotated data. Furthermore, in the case of COVID-19 diagnosis it could be relevant as data relating to patients are not completely publics.


\begin{table}[!ht]
\centering
\caption{Classification results with data augmentation.}
\tiny
\begin{tabular}{cccccc}
\hline Network & Accuracy & Sensitivity & Specificity & F$_1$ score & AUC\\
\hline
AlexNet & 74.50 $\pm$ 4.40 & 70.46 $\pm$ 6.37 & 79.05 $\pm$ 8.61 & 0.75 $\pm$ 0.04 & 0.83 $\pm$ 0.04\\\hline
GoogLeNet & 78.97 $\pm$ 3.70 & 75.95 $\pm$ 13.69 & 82.38 $\pm$ 10.53 & 0.79 $\pm$ 0.06 & 0.91 $\pm$ 0.04\\\hline
SqueezeNet & 78.52 $\pm$ 7.56 & 91.56 $\pm$ 7.63 & 63.81 $\pm$ 23.79 & 0.82 $\pm$ 0.04 & 0.90 $\pm$ 0.01\\\hline
ShuffleNet & 86.13 $\pm$ 10.16 & 83.54 $\pm$ 19.89 & 89.05 $\pm$ 5.77 & 0.86 $\pm$ 0.12 & 0.93 $\pm$ 0.06 \\\hline
ResNet-18 & 90.16 $\pm$ 2.36 & 89.45 $\pm$ 7.31 & 90.95 $\pm$ 9.29 & 0.91 $\pm$ 0.02 & 0.96 $\pm$ 0.05 \\\hline
ResNet-50 & 92.62 $\pm$ 4.19 & 91.14 $\pm$ 3.35 & 94.29 $\pm$ 5.15 & 0.93 $\pm$ 0.04 & 0.98 $\pm$ 0.01 \\\hline
ResNet-101 & 89.71 $\pm$ 10.05 & 82.28 $\pm$ 20.09 & 98.10 $\pm$ 2.18 & 0.89 $\pm$ 0.12 & 0.97 $\pm$ 0.03 \\\hline
Xception & 85.68 $\pm$ 6.76 & 90.72 $\pm$ 4.79 & 80.00 $\pm$ 19.64 & 0.87 $\pm$ 0.05 & 0.94 $\pm$ 0.04 \\\hline
Inception-v3 & 91.28 $\pm$ 8.25 & 90.30 $\pm$ 5.12 & 92.38 $\pm$ 11.98 & 0.92 $\pm$ 0.08 & 0.97 $\pm$ 0.02 \\\hline
Inception-ResNet-v2 & 86.35 $\pm$ 5.71 & 88.19 $\pm$ 6.37 & 84.29 $\pm$ 14.50 & 0.87 $\pm$ 0.05 & 0.95 $\pm$ 0.05 \\\hline
VGG-16 & 78.52 $\pm$ 10.02 & 74.68 $\pm$ 30.14 & 82.86 $\pm$ 15.91 & 0.76 $\pm$ 0.17 & 0.91 $\pm$ 0.04 \\\hline
VGG-19 & 83.22 $\pm$ 5.85 & 90.72 $\pm$ 3.19 & 74.76 $\pm$ 12.96 & 0.85 $\pm$ 0.04 & 0.90 $\pm$ 0.05 \\\hline
DenseNet-201 & 91.72 $\pm$ 6.52 & 88.61 $\pm$ 8.86 & 95.24 $\pm$ 4.36 & 0.92 $\pm$ 0.07 & 0.97 $\pm$ 0.03 \\\hline
MobileNet-v2 & 87.25 $\pm$ 10.46 & 95.78 $\pm$ 2.64 & 77.62 $\pm$ 21.63 & 0.89 $\pm$ 0.08 & 0.95 $\pm$ 0.04 \\\hline
NasNet-Mobile & 83.45 $\pm$ 7.36 & 84.81 $\pm$ 2.19 & 81.90 $\pm$ 17.46 & 0.85 $\pm$ 0.05 & 0.94 $\pm$ 0.04 \\\hline
NasNet-Large & 85.23 $\pm$ 8.25 & 79.32 $\pm$ 16.28 & 91.90 $\pm$ 5.77 & 0.84 $\pm$ 0.10 & 0.93 $\pm$ 0.05 \\\hline
\textbf{Ensemble} & 96.38 $\pm$ 4.31 & 95.95 $\pm$ 4.51 & 96.86 $\pm$ 7.03 & 0.97 $\pm$ 0.04 & 0.98 $\pm$ 0.03 \\\hline
\end{tabular}
\label{reswithau}
\end{table}

\begin{table}[!ht]
\centering
\tiny
\caption{Classification results without data augmentation.}
\begin{tabular}{cccccc}
\hline Network & Accuracy & Sensitivity & Specificity & F$_1$ score & AUC\\
\hline
AlexNet  &  86.85 $\pm$ 13.66  &  80.25 $\pm$ 22.49 & 94.29 $\pm$ 4.84 & 0.85 $\pm$ 0.16 & 0.94 $\pm$ 0.04\\\hline
GoogLeNet & 93.83 $\pm$ 6.97 & 96.71 $\pm$ 4.06 & 90.57 $\pm$ 10.53 & 0.94 $\pm$ 0.06 & 0.96 $\pm$ 0.04\\\hline
SqueezeNet & 87.52 $\pm$ 6.45 & 86.84 $\pm$ 10.11 & 88.29 $\pm$ 12.01 & 0.88 $\pm$ 0.06 & 0.94 $\pm$ 0.06\\\hline
ShuffleNet & 95.97 $\pm$ 5.09 & 95.44 $\pm$ 7.47 & 96.57 $\pm$ 2.96 & 0.96 $\pm$ 0.05 & 0.97 $\pm$ 0.03\\\hline
ResNet-18 & 95.44 $\pm$ 8.02 & 98.99 $\pm$ 1.65 & 91.43 $\pm$ 15.25 & 0.96 $\pm$ 0.07 & 0.98 $\pm$ 0.03\\\hline
ResNet-50 & 93.62 $\pm$ 6.17 & 95.57 $\pm$ 6.27 & 91.43 $\pm$ 6.06 & 0.94 $\pm$ 0.06 & 0.98 $\pm$ 0.02\\\hline
ResNet-101 & 93.29 $\pm$ 5.69 & 96.20 $\pm$ 1.79 & 90.00 $\pm$ 10.10 & 0.94 $\pm$ 0.05 & 0.98 $\pm$ 0.02\\\hline
Xception & 91.11 $\pm$ 10.14 & 89.56 $\pm$ 12.55 & 92.86 $\pm$ 7.80 & 0.91 $\pm$ 0.10 & 0.96 $\pm$ 0.03\\\hline
Inception-v3 & 93.62 $\pm$ 5.22 & 96.20 $\pm$ 0.00 & 90.71 $\pm$ 11.11 & 0.94 $\pm$ 0.07 & 0.97 $\pm$ 0.04\\\hline
Inception-ResNet-v2 & 88.59 $\pm$ 7.59 & 89.24 $\pm$ 2.69 & 87.86 $\pm$ 13.13 & 0.89 $\pm$ 0.07 & 0.96 $\pm$ 0.05\\\hline
VGG-16 & 89.26 $\pm$ 8.80 & 92.83 $\pm$ 6.24 & 85.24 $\pm$ 14.45 & 0.90 $\pm$ 0.08 & 0.96 $\pm$ 0.03\\\hline
VGG-19 & 90.16 $\pm$ 7.72 & 87.34 $\pm$ 10.36 & 93.33 $\pm$ 5.77 & 0.90 $\pm$ 0.08 & 0.97 $\pm$ 0.03\\\hline
DenseNet-201 & 96.20 $\pm$ 4.95 & 95.78 $\pm$ 5.27 & 96.67 $\pm$ 4.59 & 0.96 $\pm$ 0.05 & 0.98 $\pm$ 0.03\\\hline
MobileNet-v2 & 95.97 $\pm$ 7.18 & 96.71 $\pm$ 6.04 & 95.14 $\pm$ 8.55 & 0.96 $\pm$ 0.07 & 0.97 $\pm$ 0.05\\\hline
NasNet-Mobile & 89.26 $\pm$ 8.14 & 91.56 $\pm$ 5.12 & 86.67 $\pm$ 13.27 & 0.90 $\pm$ 0.07 & 0.95 $\pm$ 0.06\\\hline
NasNet-Large & 88.59 $\pm$ 7.59 & 90.51 $\pm$ 0.90 & 86.43 $\pm$ 17.17 & 0.90 $\pm$ 0.06 & 0.96 $\pm$ 0.03\\\hline
\textbf{Ensemble} & 96.51 $\pm$ 6.34 & 96.96 $\pm$ 4.79 & 96.00 $\pm$ 8.17 & 0.97 $\pm$ 0.06 & 0.99 $\pm$ 0.03\\\hline
\end{tabular}
\label{reswithoutau}
\end{table}

\begin{table}[!ht]
\centering
\caption{Comparison with other methods.}
\begin{tabular}{cc}
\hline Method & Accuracy\\
\hline
Zhao et al. \cite{zhao2020relation}  &  78.6 \\\hline
Bernheim et al. \cite{bernheim2020chest} & 88.0 \\ \hline
Gozes et al. \cite{gozes2020rapid} &  95.0 \\ \hline
Zheng et al \cite{zheng2020deep} & 90.1 \\ \hline
Ai et al. \cite{ai2020correlation} & 97.0 \\ \hline
Fang et al. \cite{fang2020sensitivity} & 98.0 \\ \hline
Shah et al. (Ctnet-10) \cite{shah2020diagnosis}  & 82.1 \\ \hline
Shah et al. (VGG-19)  \cite{shah2020diagnosis} & 94.5 \\ \hline
\textbf{Ensemble (data augmentation)} & 96.3\\ \hline
\textbf{Ensemble (no data augmentation)} & 96.5 \\ \hline
\end{tabular}
\label{comparison}
\end{table}









\section{Conclusions and Future Works}
\label{conc}

The challenge in COVID-19 detection is especially interesting and, not only, when the data comes from visual information. The complexity of the task is linked to different factors such as the constant increase and variation of data, given that the challenge is in full swing. In support, the convolutional neural networks give a big hand for understand the meaning of information inside the images with the consequent goal of their classification. In this context, we have proposed a framework that combines convolutional neural networks, adapted to COVID-19 detection task, through a transfer learning approach, using an ensemble criteria. The results produced certainly support the theoretical thesis. A multiple model, based on different deep neural networks, compared to a single one is a high discrimination factor. The extensive experimental phase has shown how the proposed approach is competitive, and in some cases surpassing, with respect to state of the art methods. Certainly, the main weak point concerns the computational complexity relating to learning phase, as it is known, takes a long time especially when the data to be processed grows. Future work will certainly concern the study and analysis of convolutional neural networks still unexplored for this type of problem and the application of the proposed framework to additional datasets with purpose to definitively win the challenge to COVID-19.

\section*{Acknowledgements}
Our thinking is for Alfredo Petrosino. He surely would have made a fundamental contribution to winning the battle versus COVID-19 as a man who loved challenges. Finally, we cannot be close to all the families in the world who unfortunately have lost their beloved members.



\bibliographystyle{elsarticle-num} 
\bibliography{bibliography}





\end{document}